\documentclass[conference]{IEEEtran}

\usepackage{cite}
\usepackage[cmex10]{amsmath}

\usepackage{tabularx,colortbl}
\usepackage{amsmath,amssymb,amsthm,mathrsfs,dsfont}

\usepackage{amsmath}
\usepackage{mathtools}
\usepackage{amssymb}

\usepackage{booktabs}
\usepackage{threeparttable}
\usepackage{multirow}

\usepackage{algorithm}
\usepackage{algpseudocode}

\usepackage{subfigure}
\usepackage{stfloats}

\newtheorem{proposition}{Proposition}

\newcounter{MYtempeqncnt}
\IEEEoverridecommandlockouts
\ifCLASSINFOpdf
  \usepackage[pdftex]{graphicx}
  \DeclareGraphicsExtensions{.pdf,.jpeg,.png}
\else
  \usepackage[dvips]{graphicx}
  \DeclareGraphicsExtensions{.eps}
\fi

\begin{document}

\title{{Spectrum Sharing in RF-Powered Cognitive Radio Networks using Game Theory}}

\author{\IEEEauthorblockN{Yuanye Ma$^\dag$, He (Henry) Chen$^\dag$, Zihuai Lin$^\dag$, Branka Vucetic$^\dag$, and Xu Li$^\ddag$  }
\IEEEauthorblockA{$^\dag$The University of Sydney, Sydney, Australia, Email: \{yuanye.ma, he.chen, zihuai.lin, branka.vucetic\}@sydney.edu.au\\
}
\IEEEauthorblockA{$^\ddag$Beijing Jiaotong University, Beijing, China, Email: xli@bjtu.edu.cn\\
}
\thanks{This work of Yuanye Ma was supported by China Scholarship Council and Norman I Price Supplementary Scholarship.}
}

\maketitle
\begin{abstract}
We investigate the spectrum sharing problem of a radio frequency (RF)-powered cognitive radio network, where a multi-antenna secondary user (SU) harvests energy from RF signals radiated by a primary user (PU) to boost its available energy before information transmission. In this paper, we consider that both the PU and SU are rational and self-interested. Based on whether the SU helps forward the PU's information, we develop two different operation modes for the considered network, termed as non-cooperative and cooperative modes. In the non-cooperative mode, the SU harvests energy from the PU and then use its available energy to transmit its own information without generating any interference to the primary link. In the cooperative mode, the PU employs the SU to relay its information by providing monetary incentives and the SU splits its energy for forwarding the PU's information as well as transmitting its own information. Optimization problems are respectively formulated for both operation modes, which constitute a Stackelberg game with the PU as a leader and the SU as a follower. We analyze the Stackelberg game by deriving solutions to the optimization problems and the Stackelberg Equilibrium (SE) is subsequently obtained. Simulation results show that the performance of the Stackelberg game can approach that of the centralized optimization scheme when the distance between the SU and its receiver is large enough.
\end{abstract}

\IEEEpeerreviewmaketitle

\section{Introduction}
Cooperative cognitive radio (CR) technique has been treated as a promising technology to improve the performance of CR networks \cite{6674154}. The basic idea is that the secondary user (SU) helps the primary user's (PU's) data transmission, and in return the SU can transmit its own information by utilizing the PU's spectrum. To further enhance the spectrum efficiency of cooperative CR networks, the idea of implementing multiple antennas at the secondary transmitter (ST) has been proposed and studied in \cite{manna2011cooperative,zheng2013cooperative}, which can provide additional degree of freedom by enabling the ST to relay the PU's information in addition to transmitting its own information concurrently.

The performance of cooperative CR systems may be largely constrained if the SU has limited energy to assist the PU's data transmission. Powering the cooperative CR network with radio frequency (RF) energy provides a spectrum-efficient and energy-efficient solution \cite{mohjazi2015rf}. In a RF-powered cooperative CR network, with the embedded energy harvesting component, the SU is enabled to harvest and store energy from RF signals radiated by the PU and thus has more opportunities to help the PU's data transmission. The authors in \cite{zheng2014information} proposed an information and energy cooperation scheme for CR networks, where the PU and SU are cooperative with each other and the PU can not only send information for relaying but also feed the SU with energy via RF energy transfer. Three schemes that enable information as well as energy cooperation were proposed. By assuming that the PU and SU are selfless and fully cooperative, the optimal resource allocation problems for the proposed three schemes were formulated and addressed to maximize the SU's rate subject to PU's rate and ST's power constraints \cite{zheng2014information}. However, the PU and SU can be rational and self-interested in practice. In this case, incentives should be provided by the PU to employ the SU as its information relay. The SU needs to evaluate the tradeoff between the benefits of relaying PU's information and that of transmitting its own information. To the best knowledge of the authors, there is no report in open literature that characterizes the strategic interactions between the rational PU and SU in RF-powered CR networks. This gap motivates this paper.

In this paper, we investigate the spectrum sharing of a RF-powered CR network, where the PU and the energy harvesting SU are assumed to be rational and only aim to maximize their own utilities. Game theory is employed to study the considered network since it offers a set of mathematical tools to model the complex interactions among the rational players. Note that game theory has been applied to investigate different setups of RF-powered communication networks in \cite{Chen_ISIT_2014_A,Chen_TWC_2015_Dis,Chen_ICASSP_2015_A,Ma_ICC_2015_Dis}. But none of them considered the RF-powered CR networks. \emph{The main contributions of this paper are summarized as follows:} Based on whether the SU helps the PU to relay its information, two different operation modes, named non-cooperative and cooperative modes, are identified. For both modes, optimization problems are respectively formulated, which constitute a Stackelberg game with the PU as a leader and the SU as a follower. We then analyze the Stackelberg Equilibrium (SE) of the formulated game by deriving elaborated solutions to the optimization problems. Simulation results show that the performance of the formulated Stackelberg game can approach that of the centralized optimization scheme when the distance between the SU and its receiver is large enough.

\textbf{\emph{Notations}}: Throughout this paper, we adopt the following notations. We use boldface lowercase and uppercase letters to represent vectors and matrices, respectively. $\|\cdot\|$ denotes the Frobenius norm. $(\cdot)^\dag$ denotes the conjugate transpose and $\mathbb{E}[\cdot]$ denotes the expectation.

\section{System Model}

We consider a CR network consisting of one PU pair and one SU pair. The PU pair is composed of one primary transmitter (PT) and one primary receiver (PR). The SU pair has one ST and one secondary receiver (SR). The ST is assumed to be equipped with $N \geq 2$ antennas as well as an energy harvesting component, where $N$ denotes the number of antennas. Other terminals are equipped with only single antenna. Channel reciprocity is assumed. We use $h_P$, $h_{PS}$, $\mathbf{h}_{S}$, $\mathbf{g}_{P}$, $\mathbf{g}_{S}$ to denote the channel gains between the PT and the PR, the PT and the SR, the PT and the ST, the ST and the PR, the ST and the SR, respectively. $\mathbf{h}_{S}$, $\mathbf{g}_{P}$ and $\mathbf{g}_{S}$ are assumed with the size of $N \times 1$. The duration of one transmission block is denoted by $\mathcal{T}$, over which the channel gains are assumed to be constant. Without loss of generality, we normalize the transmission duration $\mathcal{T}$ to be unity hereafter, i.e., $\mathcal{T} = 1$. 
\begin{figure}[!t]
\centering \scalebox{0.3}{\includegraphics{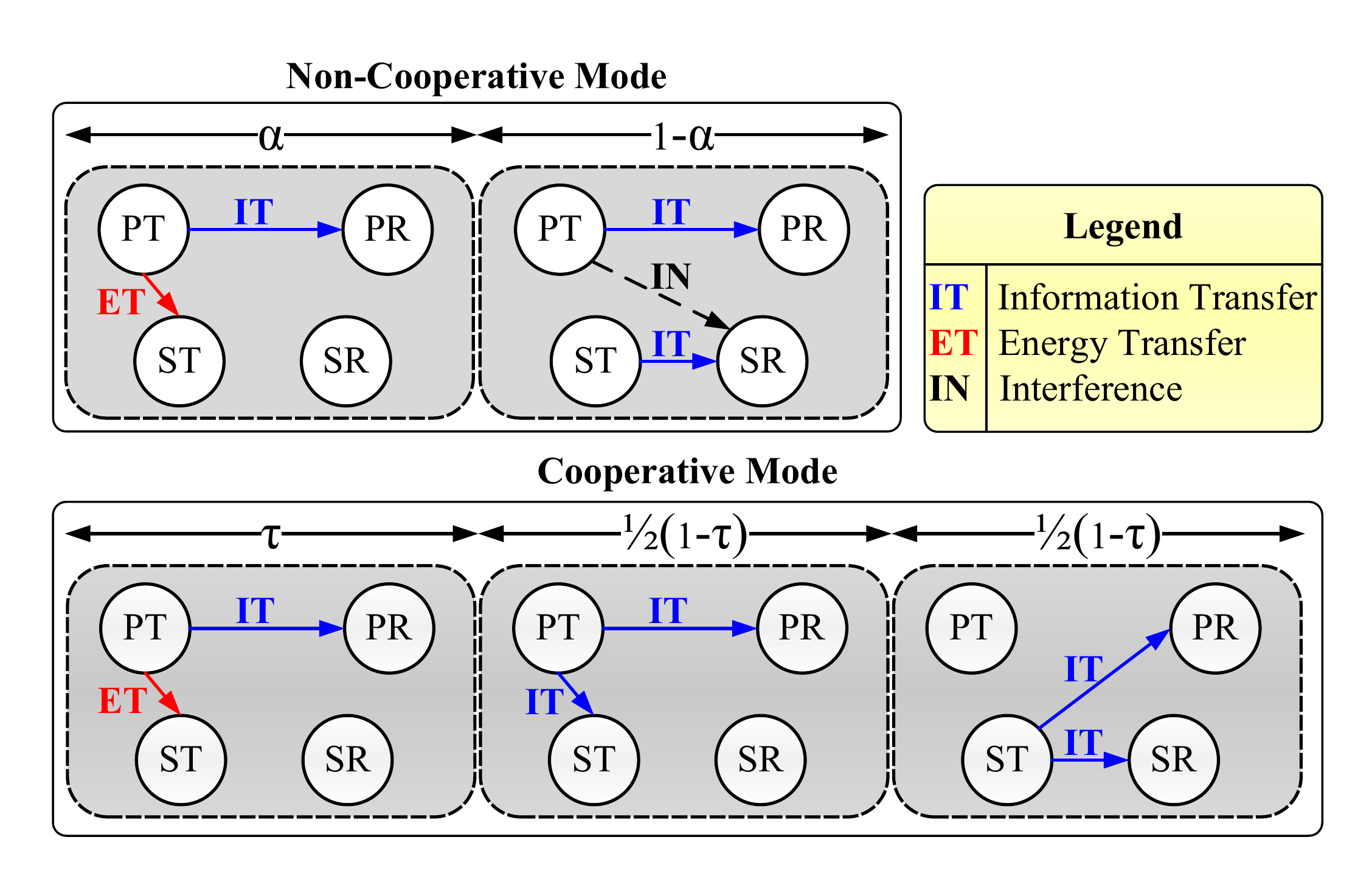}}
\caption{The non-cooperative and cooperative operation modes.}\label{cases}
\end{figure}

As illustrated in Fig. \ref{cases}, we identify two different operation modes for the considered network, named non-cooperative and cooperative modes, based on whether the SU assists the PU's information transmission. In the \textit{non-cooperative mode}, the PT transmits information to its PR using the entire transmission block. The ST first harvests energy from the PT with a duration $\alpha$, $0\leq \alpha \leq 1$, and transmits its own information to the SR during the remaining time without bringing any interference to the PR. Note that when the ST transmits to the SR, the SR has to suffer the interference from the PT. In the \textit{cooperative mode}, each transmission block is divided into three phases. The first phase is set with a fraction $\tau$, $0 \leq  \tau  < 1$, during 
which the PT transmits information to the PR and the ST harvests energy from the PU's signal. The motivation behind this energy transfer phase is that the initial energy of the SU is too small to effectively assist the PU's information transmission. The second and third phases have a same duration of $(1-\tau)/2$ by assuming that the amplify-and-forward (AF) relaying protocol is adopted at the ST. During the second phase, the PT broadcasts its information to the PR as well as the ST. During the third phase, the ST will split a portion 
$\beta$, $0 < \beta \leq 1$, of its available energy\footnote{The available energy at the ST can be the sum of its initial energy and the harvested energy from the PT.} to forward the PT's information received in the second phase. The PT will be silent during this phase, while the ST amplifies and forwards the PT's information to the PR using the portion $\beta$ of its energy. At the same time, the ST transmits its own information to the SR by using its remaining energy with the portion $1-\beta$.

To achieve an efficient spectral sharing between the PU and SU pairs, the antenna weights at the ST should be well designed in both modes. For the combination of signals (information/energy) received at the ST from the PT, the optimal weight design should be the maximum ratio combining (MRC) scheme. For the signal (information) transmitted from the ST, we first denote by $\mathbf{w}_P$ and $\mathbf{w}_S$ the transmit weight vectors with the same size $N\times 1$ that are applied to the PT's signal and the ST's signal, respectively. The transmit weight vectors $\mathbf{w}_P$ and $\mathbf{w}_S$ should be designed such that the interference caused by the PT's signal (if the ST has) to the SR and the interference caused by the ST's signal to the PR are respectively mitigated. To achieve this, there are several techniques that can be employed, such as dirty paper coding (DPC)\cite{lin2007practical} and zero forcing (ZF) \cite{spencer2004zero}. In this paper, we follow \cite{manna2011cooperative, zheng2014information} and adopt the ZF approach due to its simplicity and low complexity. With reference to \cite{manna2011cooperative}, the transmit weight vectors for ZF approach can be expressed as
\begin{equation}\label{weit2}
{\mathbf{w}}_P = \frac{\mathbf{Z}_P^\bot \mathbf{g}_P}{\left\| \mathbf{Z}_P^\bot \mathbf{g}_P \right\|}~\textrm{and}~{\mathbf{w}}_S = \frac{\mathbf{Z}_S^\bot \mathbf{g}_S}{\left\| \mathbf{Z}_S^\bot \mathbf{g}_S \right\|},
\end{equation}
where $\mathbf{Z}_P^\bot$ and $\mathbf{Z}_S^\bot$ are the projection matrices for the PR and SR, given respectively by
\begin{equation}
\mathbf{Z}_P^\bot = \left(\mathbf{I} - \mathbf{g}_S \left(\mathbf{g}_S^\dag \mathbf{g}_S\right)^{-1} \mathbf{g}_S^\dag\right),
\end{equation}
and
\begin{equation}
\mathbf{Z}_S^\bot = \left(\mathbf{I} - \mathbf{g}_P \left(\mathbf{g}_P^\dag \mathbf{g}_P\right)^{-1} \mathbf{g}_P^\dag\right).
\end{equation}

We now derive the achievable rate per unit bandwidth of the PU pair and the SU pair in both modes, respectively. 
In the non-cooperative mode, the achievable rate at the PR can be expressed as
 \begin{equation}
R_P^{NC} = \log_2 \left(1 + \frac{P |h_P|^2}{\sigma^2}\right),
\end{equation}
where $P$ is the PT transmit power and $\sigma^2$ is the noise power, which is assumed to be the same for all channels without loss of generality. The achievable rate at the SR is
\begin{equation}\label{rate_S_case2}
\begin{split}
R_S^{NC}= (1-\alpha) \log_2 \left(1 +  \gamma\left(\alpha\right)\right),
\end{split}
\end{equation}
with the signal-to-interference-plus-noise ratio (SINR)
\begin{equation}
\begin{split}
\gamma\left(\alpha\right) = \frac{ \left(\alpha \eta  \left(P \left\|\mathbf{h}_S\right\|^2  + \sigma^2 \right) + E_0\right) \left\|\mathbf{Z}_S^\bot \mathbf{g}_S \right\|^2}{(1-\alpha)\left(P \left|h_{PS}\right|^2 + \sigma^2 \right)},
\end{split}
\end{equation}
where $0<\eta<1$ is the energy conversion efficiency and $E_0$ is the initial energy at the ST. In the cooperative mode, the achievable rate at the PR can be expressed as (\ref{Rate_P_case1}) on top of next page\footnote{The details are omitted due to space limitation. We refer the interested readers to \cite{manna2011cooperative,zheng2014information}.}
\begin{figure*}[!t]
\normalsize
\setcounter{MYtempeqncnt}{\value{equation}}
\begin{equation}\label{Rate_P_case1}
\begin{split}
 R_P^{CO} =  \tau   \log_2 \left(1 + \frac{P \left| h_P \right|^2}{\sigma^2}\right)+ \frac{1-\tau}{2} \log_2 \left(1 + \frac{P|h_P|^2}{\sigma^2} +
\frac{2 P\|\mathbf{h}_S\|^2  \beta E \left(\tau\right)  \left\| \mathbf{Z}_P^\bot \mathbf{g}_P \right\|^2 }{P\|\mathbf{h}_S\|^2 (1-\tau)\sigma^2 +  2  \beta E\left(\tau \right)  \left\| \mathbf{Z}_P^\bot \mathbf{g}_P \right\|^2  \sigma^2 + (1-\tau)\sigma^4 }
\right).
\end{split}
\end{equation}
\vspace*{4pt}
\hrulefill
\end{figure*}
with
\begin{equation}
\begin{split}
E\left(\tau\right) = \tau \eta  \left(P \left\|\mathbf{h}_S\right\|^2  + \sigma^2 \right) +E_0,
\end{split}
\end{equation}
which is the ST's available energy aggregated by the harvested energy and its initial energy. The achievable rate at the PR is
\begin{equation}\label{Rate_S_case1}
\begin{split}
R_S^{CO}  = \frac{1-\tau}{2}  \log_2 \left(1 + \frac{2(1-\beta)  E\left(\tau \right)  \left\|\mathbf{Z}_S^\bot \mathbf{g}_S \right\|^2}{(1-\tau)\sigma^2} \right).
\end{split}
\end{equation}

\section{Stackelberg Game Formulation}
A Stackelberg game is formulated in this section. Recall that the PU pair and the SU pair are assumed to be rational and self-interested. We model the non-cooperative and cooperative optimizations of the PU pair and the SU pair as a Stackelberg game, in which the PU pair is modeled as the leader, as it has the priority to use the channel, and the SU pair is the follower. To formulate the Stackelberg game, we first present the optimization problems in both modes, respectively.

In the non-cooperative mode, the utility of the PU pair is a constant, which can be expressed as
\begin{equation}
\begin{split}
U_P^{NC} = &\lambda_P R_P^{NC},
\end{split}
\end{equation}
where the weight $\lambda_P >0$ is the gain per unit achievable rate of the PU pair. The SU pair can decide the duration of energy harvesting by adjusting $\alpha$ to maximize its achievable rate. The utility function of the SU pair can be written as
\begin{equation}\label{}
\begin{split}
U_S^{NC}(\alpha) =& \lambda_S R_S^{NC}\left(\alpha\right),
\end{split}
\end{equation}
where the weight $\lambda_S >0$ is the gain per unit achievable rate of the SU pair. The optimization problem of the SU in the non-cooperative mode can thus be given by
\begin{equation}\label{Eq.SO2}
\begin{split}
\max_{\alpha} ~ U_S^{NC}(\alpha),~\textrm{s.t.}~ 0 \leq  \alpha \leq 1,
\end{split}
\end{equation}
whose optimal solution is denoted by $\alpha^\star$.

In the cooperative mode, we consider a pricing-based resource allocation to model the utilities of the PU and SU pairs as monetary incentives \cite{kang2012price,xu2014pricing} should be provided by the PU to hire the SU as its information relay. We assume that the PT would like to provide a price $\mu > 0$ per unit energy to hire the SU as its relay. Besides, the PT (leader) could decide the duration of $\tau$ to control the time allocation of the three phases. The ST (follower) will choose a splitting ratio $\beta$ to maximize its own utility based on the given price and time. Then, the utility function of the PU pair can be expressed by
\begin{equation}\label{Up1}
\begin{split}
U_P^{CO}(\tau, \mu ) = \lambda_P R_P^{CO}\left(\tau\right) - \mu \beta E\left(\tau\right).
\end{split}
\end{equation}
 The utility function of the SU pair thus can be written by
\begin{equation}
\begin{split}
U_S^{CO}(\beta) =& \lambda_S R_S^{CO}\left(\beta\right) +  \mu \beta  E\left(\tau\right).
\end{split}
\end{equation}
Then, the optimization problems for the PU pair and the SU pair in the cooperative mode can be respectively formulated~as
\begin{equation}\label{Eq.PO}
\begin{split}
\max_{\tau, \mu} ~ U_P^{CO}(\tau, \mu ),~\textrm{s.t.}~0\leq \tau < 1,~\mu > 0,
\end{split}
\end{equation}
and
\begin{equation}\label{Eq.SO}
\begin{split}
\max_{\beta} ~ U_S^{CO}(\beta),~\textrm{s.t.}~ 0 <  \beta \leq 1,
    \end{split}
\end{equation}
whose optimal solutions are denoted by $\left(\tau^\star, \mu^\star\right)$ and $\beta^\star$.

A Stackelberg game thus can be formulated by putting the optimization problems given in (\ref{Eq.SO2}), (\ref{Eq.PO}) and (\ref{Eq.SO}) together. Once a Stackelberg game is formulated, the next step is to find the Stackelberg Equilibrium (SE) point from which neither the leader nor the follower has incentives to deviate. It is worth noting that as the leader, the PU pair can decide the strategy to fulfil either the non-cooperative mode or the cooperative mode by comparing the maximized utilities of these two modes. If $U_P^{CO} (\tau^\star, \mu^\star) > U_P^{NC}$ and $U_S^{CO}(\beta^\star) > U_S^{NC}(\alpha^\star)$, 
the PU pair will choose the cooperative mode and the SE of the game will be $\left(\tau^\star, \mu^\star, \beta^\star\right)$; otherwise, the PU pair will choose the non-cooperative mode and the SE will be $\alpha^\star$.

\section{Stackelberg Game Analysis}
In this section, we will derive the SE by solving the optimization problems (\ref{Eq.SO2}), (\ref{Eq.PO}) and (\ref{Eq.SO}). The optimal solution $\alpha^\star$ to the optimization problem (\ref{Eq.SO2}) can be readily obtained by conducting a one-dimensional exhausted search, i.e.,
\begin{equation}\label{alpha}
\begin{split}
\alpha^\star =  \arg  \max_{ \alpha \in [0,1]} U_S^{NC}\left(\alpha\right).
    \end{split}
\end{equation}
We thus focus on the optimization problems (\ref{Eq.PO}) and (\ref{Eq.SO}) in the cooperative mode in the following. 

First, we solve the optimization problem (\ref{Eq.SO}) with given $\mu$ and $\tau$ as shown in the following proposition.
\begin{proposition}\label{Pro1}
Given $\mu$ and $\tau$, the optimization problem (\ref{Eq.SO}) has a valid solution conditioned on\footnote{The condition in (\ref{c1}) is to ensure that the optimal value of $\beta$ is large than zero. Otherwise, the SU may reject to cooperate.}
\begin{equation}\label{c1}
\begin{split}
\mu > \frac{XY}{X+1},
\end{split}
\end{equation}
and the optimal solution $\beta^\star$ to the problem (\ref{Eq.SO}) can be expressed by
\begin{equation}\label{beta}
\beta^\star =
\begin{cases}
\beta^* &\mbox{if $ \frac{XY}{X+1} < \mu \leq XY$},\\
1  &\mbox{if $\mu  >  XY $},
\end{cases}
\end{equation}
where
\begin{equation}\label{XY}
\begin{split}
X = \frac{2  E\left(\tau \right) \left\|\mathbf{Z}_S^\bot \mathbf{g}_S \right\|^2}{(1-\tau)\sigma^2},~Y = \frac{\lambda_S (1-\tau)  }{2  E\left(\tau \right) \ln 2},
\end{split}
\end{equation}
and
\begin{equation}
\begin{split}
\beta^* = \frac{1}{X} - \frac{Y}{\mu} + 1.
\end{split}
\end{equation}
\end{proposition}
\begin{proof}
The proof is omitted due to space limitation.
\end{proof}

Then, we can replace $\beta$ by its optimal value $\beta^\star$ in $U_P^{CO}(\tau, \mu)$ to solve the optimal $\mu^\star$ and $\tau^\star$. As $\beta^\star$ has two expressions conditioned on the different intervals of $\mu$, we thus can write the expression of $U_P^{CO}\left(  \tau, \mu, \beta^\star \right)$~as
\begin{equation}
U_P^{CO}\left(  \tau, \mu, \beta^\star \right) =
\begin{cases}
     U_P^{CO} \left(  \tau, \mu, \beta^* \right) &\mbox{if $\frac{XY}{X+1} < \mu \leq  XY $},\\
     U_P^{CO} \left(  \tau, \mu, 1 \right) &\mbox{if $\mu >  XY $},
\end{cases}
\end{equation}
where $U_P^{CO} \left(  \tau, \mu, \beta^* \right)$ is given in (\ref{Eq.UP_beta}) on top of next page
\begin{figure*}[!t]
\normalsize
\setcounter{MYtempeqncnt}{\value{equation}}
\begin{equation}\label{Eq.UP_beta}
\begin{split}
U_P^{CO}(\tau, \mu , \beta^*) = \lambda_P  \tau   \log_2A + \lambda_P   \frac{1-\tau}{2} \log_2 \left(A +  \frac{ \mu B \left(\frac{1}{X} +1\right) - BY  }{\mu C + \mu D  \left(\frac{1}{X} +1\right) - DY }\right) - \mu \left(\frac{1}{X} +1\right) E\left(\tau\right) + Y E\left(\tau\right).
\end{split}
\end{equation}
\vspace*{4pt}
\hrulefill
\end{figure*}
and
\begin{equation}\label{UP01}
\begin{split}
U_P^{CO}(\tau, \mu,  1) = &  \lambda_P \tau   \log_2A  - \mu  E\left(\tau\right)\\
&+ \lambda_P  \frac{1-\tau}{2}  \log_2 \left(A +  \frac{B}{C+D }\right),
\end{split}
\end{equation}
with $A > 0$, $B > 0$, $C> 0$ and $D >0$, respectively, given by
\begin{equation}
\begin{split}
A  = 1+ \frac{P \left| h_P \right|^2}{\sigma^2},
\end{split}
\end{equation}
\begin{equation}
\begin{split}
B  = 2 P\|\mathbf{h}_S\|^2  E\left(\tau\right)  \left\| \mathbf{Z}_P^\bot \mathbf{g}_P \right\|^2,
\end{split}
\end{equation}
\begin{equation}
\begin{split}
C  = P\|\mathbf{h}_S\|^2 (1-\tau)\sigma^2 +  (1-\tau)\sigma^4,
\end{split}
\end{equation}
\begin{equation}
\begin{split}
D =   2  E\left(\tau\right)  \left\| \mathbf{Z}_P^\bot \mathbf{g}_P \right\|^2  \sigma^2.
\end{split}
\end{equation}
The maximization problem (\ref{Eq.PO}) is then updated as
\begin{equation}\label{Eq.PO2}
\begin{split}
\max_{\tau, \mu } ~ U_P^{CO}\left(\tau, \mu, \beta^\star \right),~\textrm{s.t.}~  0\leq \tau < 1,~\mu > \frac{XY}{X+1}.
\end{split}
\end{equation}
When $\mu > \frac{XY}{X+1}$, since $\beta^\star$ is a continuous function of $\mu$, $U_P\left(  \tau, \mu, \beta^\star \right)$ is a continuous function of $\mu$ as well. To solve the problem (\ref{Eq.PO2}), we first find the optimal relationship between $\tau$ and $\mu$ by expressing $\mu$ as a function of $\tau$.  By regarding $\tau$ as a constant first, we can have the following maximization problem,
\begin{equation}\label{Eq.PO3}
\begin{split}
\max_{\mu } ~ U_P^{CO}\left(\tau, \mu, \beta^\star \right),~\textrm{s.t.}~\mu > \frac{XY}{X+1}.
\end{split}
\end{equation}
We use $\widetilde{\mu}(\tau)$ to denote the optimal solution to the problem (\ref{Eq.PO3}), which can be obtained in the following proposition.
\begin{proposition}\label{Pro2}
Given $0 \leq \tau < 1$, the optimization problem (\ref{Eq.PO3}) has a valid solution conditioned on
\begin{equation}\label{c2}
\begin{split}
\mu^* > \frac{XY}{X+1},
\end{split}
\end{equation}
and the optimal solution $\widetilde{\mu}(\tau)$ to the problem (\ref{Eq.PO3}) can be expressed by
\begin{equation}\label{mu}
\widetilde{\mu} (\tau) =
\begin{cases}
     \mu^*  &\mbox{if $ \frac{XY}{X+1} < \mu^* \leq  XY $},\\
    XY &\mbox{if $ \mu^* >  XY $},
\end{cases}
\end{equation}
where $\mu^*$ is given by (\ref{mu_express}) on the top of next page.
\end{proposition}
\begin{proof}
The proof is omitted due to space limitation.
\end{proof}

\begin{figure*}[!t]
\vspace*{4pt}
\normalsize
\setcounter{MYtempeqncnt}{\value{equation}}
\begin{equation}\label{mu_express}
\begin{split}
\mu^* =& \frac{1}{2 A\left(C + D\left(\frac{1}{X} + 1\right)\right)^2 +  2 B \left(\frac{1}{X} + 1\right) \left(C + D\left(\frac{1}{X} +1\right)\right)}\left(2 A D Y\left(C + D\left(\frac{1}{X} +1\right)\right) + B Y \left(C + 2D\left(\frac{1}{X} +1\right)\right) \right.\\
 & \left. +\sqrt{B^2 C^2 Y^2 + \frac{2 \lambda_P (1-\tau)  B C Y}{\left(\frac{1}{X} + 1\right) E\left(\tau\right) \ln2} \left( A\left(C + D\left(\frac{1}{X} + 1\right)\right)^2 + B \left(\frac{1}{X} + 1\right) \left(C + D\left(\frac{1}{X} +1\right)\right) \right) }\right)
\end{split}
\end{equation}
\hrulefill
\vspace*{4pt}
\end{figure*}

Finally, by conducting a one-dimensional exhausted search, we can obtain
\begin{equation}\label{tau}
\begin{split}
\tau^\star = \arg \max_{\tau \in [0,1)} U_P^{CO}\left(  \tau, \widetilde{\mu}(\tau), \beta^\star \right),
\end{split}
\end{equation}
and $\mu^\star = \widetilde{\mu}\left(\tau^\star\right)$. We then have obtained the SE of the formulated Stackelberg game by combining (\ref{alpha}), (\ref{beta}), (\ref{mu}) and (\ref{tau}). Note that if the conditions in (\ref{c1}) and (\ref{c2}) are not satisfied, then the SE will be degraded to $\alpha^*$.

\section{Simulation Results}
In this section, we present the performance results of the formulated game via computer simulation.  We assume that all the channels experience quasi-static flat Rayleigh fading and adopt a distance-dependent pass loss model such as $L_d = 10^{-3}(d)^{-\varphi}$, where $d$ denotes the distance between arbitrary two nodes and $\varphi$ is the pass loss factor. Using the same setting in \cite{zheng2014information}, unless otherwise specified, we consider the distances from the ST to all the other terminals are $1$m, while the distance from the PT to the PR is $2$m and $\varphi$ is chosen as $3.5$. The transmit power of the PT is $30$dB. The initial energy $E_0$ at the ST is $10$dB. The noise power $\sigma^2$ is normalized to unity, i.e., $\sigma^2 = 1$. The weights $\lambda_P$ and $\lambda_S$ are set to be the same equal to $100$ and the energy conversion efficiency $\eta$ is $0.5$. In the sequel, all the shown results are averaged over $10^4$ channel realizations.

\begin{figure}[t]
\centering
 \subfigure[The utility of the PU pair.]
  {\scalebox{0.33}{\includegraphics {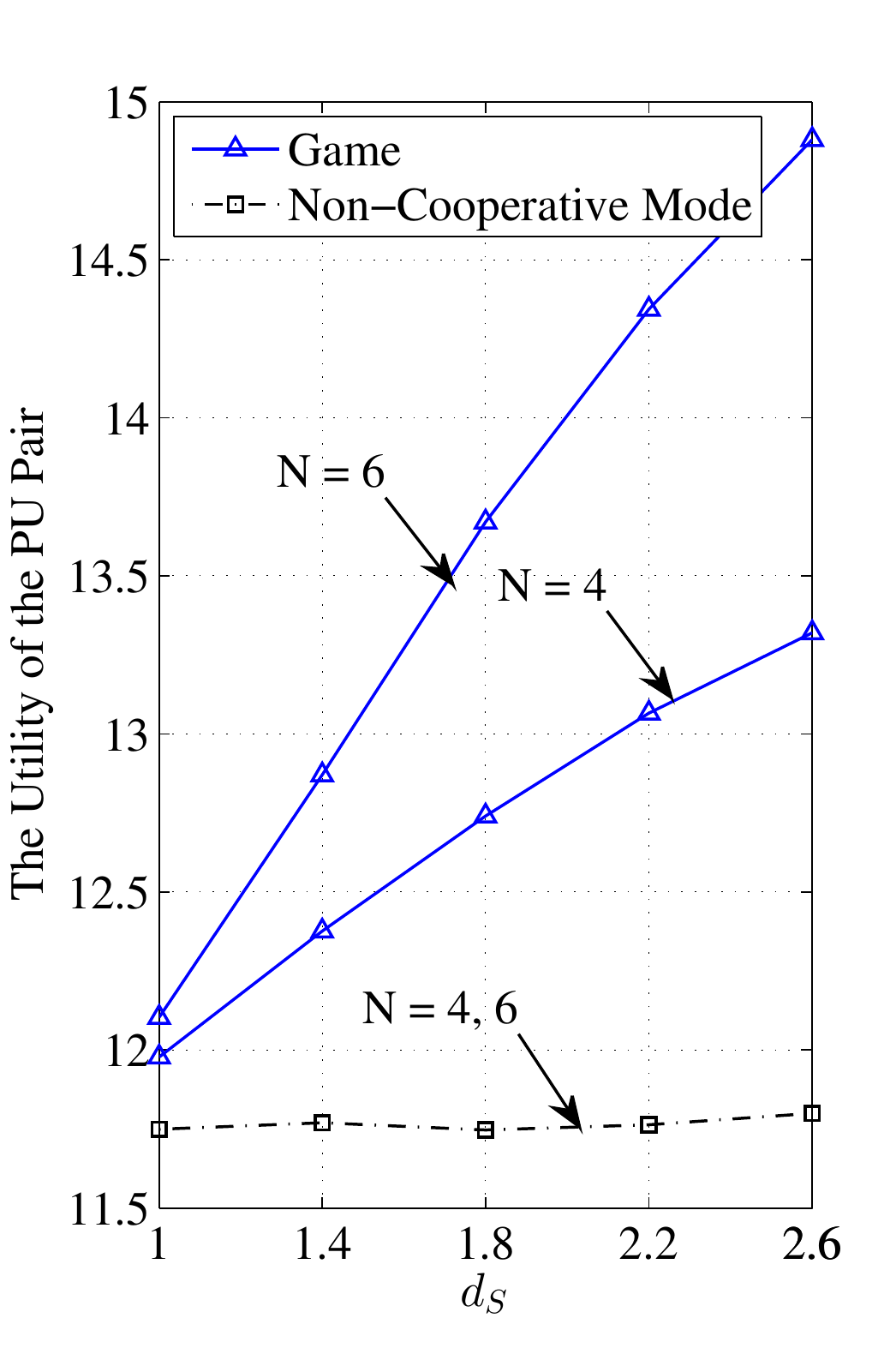}\label{fig2a}}}
 \subfigure[The utility of the SU pair.]
  {\scalebox{0.33}{\includegraphics {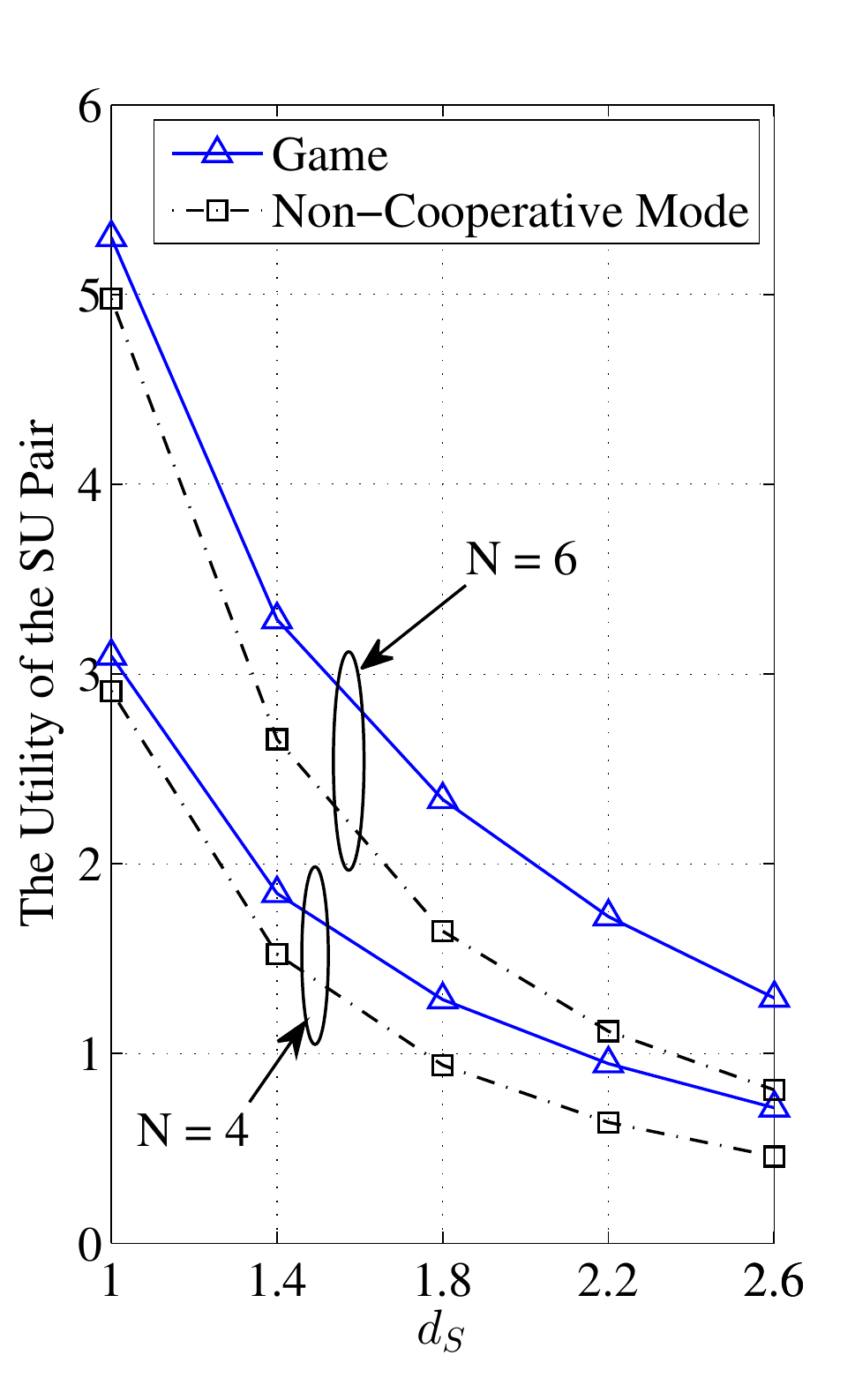}\label{fig2b}}}
 \caption{The effects of changing the distance between the ST and the SR, i.e., $d_S$, on the utilities of the PU and SU pairs, respectively.}
\label{fig2}
\end{figure}
Fig. \ref{fig2} depicts the effects of changing the distance between the ST and the SR, which is denoted by $d_S$, on the utilities of the PU and SU pairs, respectively. It can be observed in Fig. \ref{fig2a} that with the increasing of $d_S$, the PU pair's utility is improved. This is because, when $d_S$ increases, the SU pair prefers to sell a portion of its energy to assist the PU's data transmission, instead of using the energy to transmit its own information and obtain a marginal achievable rate. Conversely, it is shown in Fig. \ref{fig2b} that the SU pair's utility decreases with the increasing of $d_S$. This is because compared with the gained revenue from selling energy, the reduction of the achievable rate due to the worse channel condition dominates the performance of the SU pair's utility. Furthermore, we can observe from both Fig. \ref{fig2a} and Fig. \ref{fig2b} that the performance of the game always outperforms that of the non-cooperative mode. In addition, the number of antennas at the ST $N = 6$ brings more advantages to both the PU and SU pairs than that $N = 4$ due to the beamforming. Note that changing the number of antennas at the ST will not affect the PU pair's utility in the non-cooperative mode as shown in Fig.~\ref{fig2a}.

\begin{figure}[!t]
\centering \scalebox{0.45}{\includegraphics{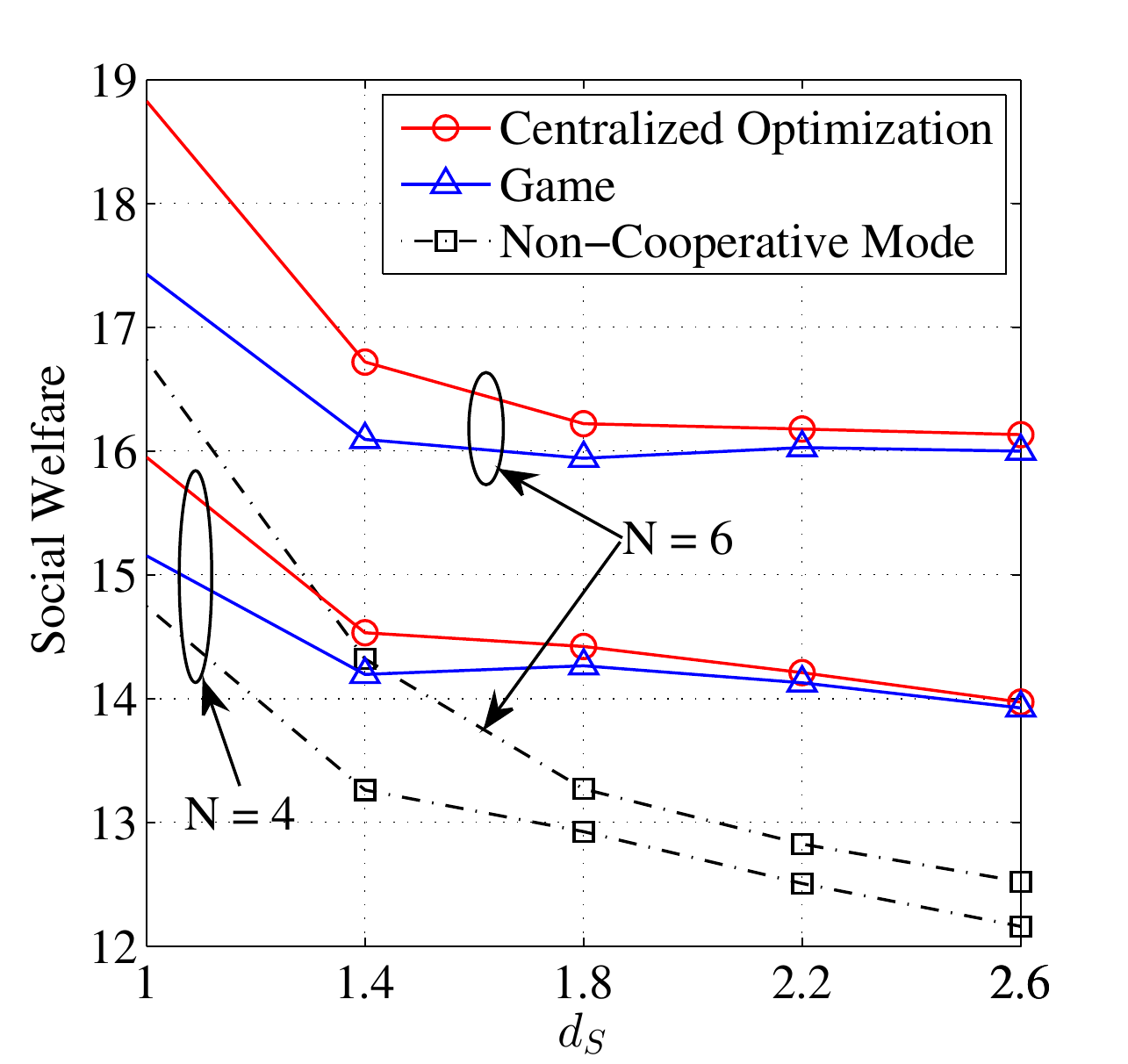}}
\caption{The social welfare performance versus the distance between the ST and the SR ($d_S$) with the number of antennas $N = 4$ and $N = 6$ at the~ST.}\label{fig3}
\end{figure}
Fig. \ref{fig3} illustrates the social welfare performance versus the distance between the ST and the SR, i.e., $d_S$. The social welfare is defined as the aggregation of both PU and SU pairs' utilities. We choose a centralized optimization method referring from \cite{zheng2014information} as a benchmark to compare with the performance of the game. The centralized optimization approach adopted herein is conducted by greedy search to maximize the social welfare. It is shown in Fig. \ref{fig3} that with the increasing of $d_S$, the performance of the game achieves a better performance than that of the non-cooperative mode, and approaches to the maximum social welfare optimized by the centralized method. This is because, when $d_S$ is small, the SU pair experiences a good channel between the ST and the SR. It will selfishly use the majority of its energy (harvested and initially has) to transmit its own information. When $d_S$ is larger, selling a portion of its energy to assist the PU pair's data transmission becomes a better choice to the SU pair, and thus the SU pair would like to help relay the PU pair's data transmission. Then a good social welfare performance can be achieved due to the cooperation. Besides, it can be observed that the social welfare performance with $N = 6$ always outperforms that with $N = 4$, which benefits from the~beamforming.

\section{Conclusion}

In conclusion, we formulated a Stackelberg game to characterize the spectrum sharing of a RF-powered CR network with rational and self-interested PU and SU pairs, where the PU pair is the leader and the energy harvesting SU pair is the follower. We modeled the utility functions and formulated the corresponding optimization problems for both non-cooperative and cooperative modes of the considered network. By solving the optimization problems of both modes, we obtained the SE of the game. In simulations, it is shown that with the increase of the distance between the ST and the SR, the performance of the game can approach that of the centralized optimization~scheme.

\bibliographystyle{IEEEtran}
\bibliography{endnoteEHCNR}

\end{document}